\documentclass[12pt]{article}
\begin{document}

\addtolength{\baselineskip}{0.5\baselineskip}

\title{\textbf{Exploring the Harmony between Theory and Computation \\
- Toward a unified electronic structure theory}}
\author{Liqiang Wei\\
Chemical Physics Research Institute~\footnote{The original version
of this paper was finished in the early of 2002 and can be found in arXiv: physics/0307156.}\\
Abington, MA 02351}

\maketitle

\begin{abstract}
\vspace{0.05in}

The physical aspect of a general perturbation theory is explored.
Its role as a physical principle for understanding the interaction
among matter with different levels of hierarchy is appreciated. It
is shown that the generic perturbation theory can not only be used
for understanding various electronic phenomena including the
nature of chemical bonds but also serve as a $\it{unified}$ theme
for developing $\it{general}$ electronic structure theories and
calculation schemes. In particular, a $\it{standard}$ electron
correlation approach is suggested and established according to
this law.
\end{abstract}

\vspace{0.35in}

$\bf{Keywords}$: unification; harmony; hierarchy; quantum
mechanics; quantum many-body theory; interaction; perturbation
theory; variational method; energy-scale principle; quantum
chemistry; electronic structure; molecular orbital; electron
correlation; multireferences; single-particle Green function;
Dyson equation; correlated molecular orbital; density-functional
theory; Kohn-Sham equation; time-dependent density-functional
theory; basis set; pseudopotential; QM/MM; intermolecular forces


\vspace{0.65in}

Perturbation theory is regarded as one of two major approaches for
approximately solving quantum many-body problems. However, its
deeper physical aspect is far more than it is at present being used
just as a mathematical apparatus for solving the complicated issues.
All the fundamental laws in physics are variational in nature,
including the Schr$\ddot{o}$dinger equation in quantum mechanics.
Nevertheless, the perturbation theory provides a basic principle
that governs how matter with different levels of hierarchy
interacts. In fact, a general perturbation method itself contains
two ingredients. On one hand, the degenerate or near-degenerate
situation is $\it{not}$ a perturbation at all but actually
constitutes a strong physical interaction. On the other hand, the
non-degenerate case is a real perturbation in the common sense. We
believe that, it is this physical mixing with equal or near energies
that governs the interaction among matter with different levels of
hierarchy. Of course, it is also an elementary physical law based on
which a $\it{unified}$ chemical bond theory can be built.

Electrons are quantum mechanical entities which possess
wave-particle duality. The binding process of the electrons
associated with some atoms, or equivalently, the interaction of
atomic orbitals for the formation of a molecule, can be regarded as
a wave interference phenomenon. The interaction of
$\it{intra}$-atomic orbitals with the same or near energies is the
Pauling's hybridization process, which determines the spatial
orientation of chemical bonds, while the interaction of
$\it{inter}$-atomic orbitals with the same energy or near energies
decides the actual formation of chemical bonds. These are the nature
of chemical bonds~\cite{pauling,ruedenberg1}. The immediate benefit
for recognizing this near energy principle in the determination of
the chemical bonds is that it gives a better understanding of many
very important structural concepts such as multi-center chemical
bonds, multiple chemical bonds, resonance structure, Walsh diagrams,
and avoided crossing. It can incorporate these different concepts
into $\it{single}$ qualitative theoretical
framework~\cite{pauling,lipscomb,harris,hoffmann,wei0,kauzmann,cotton}.

Furthermore, in addition to serving as a fundamental physical law
for understanding how matter interacts, the generic perturbation
approach can provide a $\it{unique}$ and most powerful
mathematical device for quantitatively investigating the
electronic structure of molecules including molecular materials
and biomolecules~\cite{wei0}. We are going to have a
$\it{harmony}$ between theory and computation.

$\underline{\it{Energy\ scale\ principle\ in\ the\ Rayleigh-Ritz\
variational\ approach}}$

The Rayleigh-Ritz variational method is most commonly used for
solving eigenvalue problem in quantum mechanics. Its relation to
the general perturbation theory has been the subject of analysis
for many decades, which includes the study of the complicated
issues concerning the perturbation theory for the linear
operators~\cite{slater,kato,marcus}. However, the utilization of
this relation, or its role as a guidance, in establishing the
electronic structure calculation schemes has not been fully
explored or completed yet. A related $\it{universal}$ formalism
for the quantum many-body systems is still lacking, which is of
paramount importance for our investigation~\cite{kirao}. At a
first glance, as long as the reference Hamiltonian which produces
the basis functions is made as close as possible to the full
Hamiltonian, then the dimension for the Rayleigh-Ritz variational
expansion will be made as small as possible. In addition, if the
basis functions have the closest energies of the reference
Hamiltonian, then they will have the strongest mixing and make the
greatest contribution to the combined states; while the others
with larger energy differences will have smaller or even
negligible contributions. These are the situations we
qualitatively discussed above for the elementary perturbation
theory. We term this as the $\it{energy\ scale\ principle}$ in the
Rayleigh-Ritz variational approach.

 (a) $\underline{\it{Molecular\ fragmentation\ and\ basis\ set\ construction}}$

The basis set approach is a most popular and natural way for
solving the single particle equations such as the Hartree-Fock
equation for a molecular system. Physically, it reflects a
composite relation between the molecule and its constituent atoms.
To have an overall accurate electronic structure calculation, the
first necessary step is to get the reliable and converged
molecular orbitals~\cite{wei4}.

However, since the current basis functions like contracted
Gaussians which are most commonly used are primarily a reflection
of electrons in single atoms in the molecule, it leaves the
perturbed part of the molecular Fock operator very large. That is
why the polarization functions, including some expanded ones such
as the correlation consistent basis sets, have to be introduced to
get good computation results~\cite{pople1,dunning}. Nevertheless,
the $O(N^{4})$ scaling, where $N$ is the number of basis
functions, has become a major bottleneck in quantum chemistry
calculation, especially for the large systems.

To overcome this difficulty, the energy-scale principle described
above can be helpful. If we construct the basis functions which
are the reflection of molecular fragments so that the
corresponding reference Hamiltonian is as close as possible to the
whole molecular Fock operator, then the dimension of basis set
expansion can be made as small as possible. This is going to be a
challenge work but will be mathematical in nature. The basis set
superposition effects (BSSE) is an example~\cite{kestner}.

 A similar situation occurs in the quantum molecular scattering
calculation, where the channels are used as the basis functions
for solving the Schr$\ddot{o}$dinger equation, or its integral
form, Lippmann-Schwinger equation with proper boundary conditions.
Since there are often very large differences between the channels
and the scattering waves for the whole reactive system in the
interaction regions, the dimension for their expansion is
particularly large. This causes the quantum scattering calculation
to be prohibitively expensive for all but the smallest systems.
The ideas from the perturbation theory can obviously be utilized
for remedying this deficiency~\cite{wei2}.

(b)${\underline{\it{General\ multireference\ electronic\
structure\ theory}}}$

To get a final accurate solution to the Schr$\ddot{o}$dinger
equation for the many-electron system with a non-separable
two-body Coulomb potential, it is most likely that we have to go
beyond the $\it{single}$ particle
 description~\cite{pople2,karplus1,bartlett,schaefer}.
Mathematically, the full configuration interaction (FCI) gives
exact answers~\cite{shavitt}. However, it is computationally
prohibitive and possibly will never be strictly realized. The
energy scale principle described above can also be applied in this
$\it{configuration}$ level. A general electronic structure theory
must be multiconfigurational or multireferential in
nature~\cite{goddard1,rudenberg2,peyerimhoff,roos,freed,werner,gordon,davidson1}.
First, there exists a strong configuration mixing, for example, at
transition states, for excited states, and for multiple chemical
bonds. The concept of exciton introduced in solid state physics
also belongs to this case~\cite{wei0,wei3,jortner}. Second, the
degenerate configurations are often the case for the stable
open-shell systems. Third, if we want to treat the ground state
and the excited states simultaneously, we have to include the
corresponding reference states in the same model space. Finally,
the separation of correlation into the static and dynamic parts,
which corresponds to the near degenerate and the perturbed
situations, really has chemical structure signature. Therefore,
among all the correlation approaches developed so far for the
electronic structure, the type of multiconfigurational
self-consistent fields (MCSCF) with perturbation or
coupled-cluster expansion corrections should be the most
appropriate one and works in the right direction. To solve the
issues such as proper selection of the configurations for the
model space, the efficient treatment of the dynamic correlation,
and the avoidance of the intruder states, we need not only a
mastery of current quantum many-body theories but also their
further development~\cite{fetter,lindgren}.

The density functional theory $(DFT)$ is one of the most powerful
computational schemes in the study of the electronic structures for
the molecules and solids~\cite{kohn,parr}. However, the importance
and necessity of the separation of the correlation into a static
part and a dynamic one is also indicated in its calculation of the
highly charged ions and in its treatment of the transition states
for reactions~\cite{jursic,davidson2}. Even though the
time-dependent $\it{DFT}$ has been developed to address the excited
state issues, it seems that the overall density functional theories
are still within the framework of quantum mechanics based on the
state vectors and operators~\cite{tddft}. The density is one of the
most fundamental physical quantities which is used for specifying a
system.

One most $\it{generic}$ quantum many-body approach has been
suggested and established within the perturbation
theory~\cite{wei4}. Since the molecular orbitals define a
reference Hamiltonian for the interacting many-electron
system~\cite{szabo}, it is crucial to choose the appropriate ones
to make the reference Hamiltonian as close as possible to the
whole Hamiltonian. In this way, the computational burden relevant
to the correlations will be alleviated. Nevertheless, the
correlated molecular orbitals, determined from the Dyson equation,
provide this candidacy. It is a most general energy eignequation
for generating the molecular orbitals, covering the Hartree-Fock
and Kohn-Sham equations as special cases. In practice, it gives
the best $\it{single}$-particle properties such as electron
affinities and electron ionization potentials. Meanwhile, the
one-particle Green function method can go beyond the
single-determinant level and study the $\it{N}$-body properties
such as the transition states and total energies including
excitation energies related to the configuration mixing. Since the
self-energy operator can be approached in a systematic way, the
multireference perturbation theory, combined with the
single-particle Green function formalism, will henceforth furnish
a $\it{most}$ powerful approach for studying the static and
nonstatic correlations of the interacting electrons~\cite{wei4}.

(c)${\underline{\it{General\ pseudopotential\ theory}}}$

The concepts of effective core potentials (ECP), pseudopotentials,
or model potentials (MP) are those of the most significant
developments in the fields of the electronic structure for
molecular and solid state systems~\cite{goddard2,hamann,huzinaga}.
They treat valence electrons only, leaving the core electrons and
the nucleus as a whole charge entity and therefore reducing the
number of electrons as well as the corresponding overall size of
the basis sets employed for the computation. It is important when
we study the electronic structure for the large molecules or the
inorganic molecules containing heavy elements. A type of
pseudopotentials most commonly used for the solid state
calculations is the so-called norm conserving
pseudopotentials~\cite{hamann}. In addition to having the same
energies for the valence states, their pseudo valence
wavefunctions are equivalent to the wavefunctions for the valence
states obtained from the full electron calculations outside a
cutoff radius. The pseudopotentials constructed in this manner
share the same scattering properties as those of the full
potentials over the energy range of the valence states. The
practical implementation of the various pseudopotentials has also
demonstrated the importance of choosing a correct size of the core
or the range of the valence electrons for the accurate
pseudopotential computation in order that the core-valence
correlation or core polarization can be neglected. Obviously, the
physics behind this separation of the valence and core states is
the energy scale principle we described above applied in the level
of $\it{atomic\ orbitals}$. After realizing this principle,
however, we might establish a more $\it{general}$ pseudopotential
theory. We are planning to do so within the framework of a
perturbation theory so that the most flexible and accurate
pseudopotentials or effective core potentials can be obtained.
They can be used in different chemical environments and work for
both ground and excited state problems. The final goal is to make
the effective core potentials to be a routine for the calculation
of the electronic structures for the large molecules, the
inorganic molecules containing heavy elements, and the solid state
systems.

(d)${\underline{\it{Molecular\ fragmentation\ and\ combined\
quantum\ mechanical\ and}}}$\newline ${\underline{\it{molecular\
mechanical\ (QM/MM)\ approach\ for\ electronic\ structure\ of}}}$
\newline ${\underline{\it{large\ molecules}}}$

A combined QM/MM approach has become very popular in recent years
in the investigation of, for example, the chemical reactions in
solutions and in enzymes~\cite{warshel}. The basic consideration
is that treating a full collection of electrons for the whole
system explicitly is not only unrealistic but also unnecessary. In
the first place, the electronic charge redistribution induced by a
chemical reaction is very often limited to a small region due to
the length scale issues such as a finite range of interaction or
natural charge distribution. Second, the quantum exchange effect
for the electrons is finite range, and there is no exchange
interaction among the electrons with a long
distance~\cite{karplus2,keiji,gao,friesner,jorgensen}. This
permits a partition of the whole system into an active part and an
inactive one without any charge redistribution. The former has to
be described quantum mechanically since it possibly involves bond
breaking and making, while the latter can be described by
molecular mechanics because it merely serves as a classical
electrostatic environment for the active site~\cite{allinger}.
This combined QM/MM description has shown remarkable successes in
studying the electronic structure and reactivity of large
molecules. However, challenges remain. One of the major obstacles
for the applications is in the proper treatment of the boundary
region where the cut has to be for a covalent bond. Currently,
there are two approaches to this problem. The one introducing link
atoms along the boundary is severely limited and cannot be applied
to treat a large variety of different chemical systems. In
addition, it artificially brings additional forces into the system
and therefore complicates the problem. The other kind like local
self-consistent field methods seems reasonable but it is still
more empirical. In order to utilize this kind of combined QM/MM
methods for investigating the electronic structure and molecular
dynamics in a larger domain of fields, we need to develop a more
generic ab initio approach. We believe that the energy scale
principle discussed above can play a key role here. It is not only
the principle according to which the atomic orbitals including
valence ones interact along the boundary but also the law based on
which a $\it{systematic}$ approach for constructing the correct
charge distribution or the force fields along the boundary can be
established. This is also the key for a more sophisticated or
finer treatment of the quantum region including its electron
correlation.

In summary, the energy scale principle for the hierarchy of
interacting matter is identified. Not only can it be utilized as a
$\it{universal}$ law for understanding how matter interacts at
different levels but also the relevant perturbation formalisms,
including the ones pertaining to the Rayleigh-Ritz variational
expansion, can serve as the foundation for building various and
most powerful electronic structure calculation schemes even for
the large molecular systems~\cite{wei0}. In particular, a standard
electron correlation approach or a quantum many-particle theory in
general have thereby been established~\cite{wei4}. Obviously, they
can also be employed to develop $\underline{\it{a\ generic\
theory\ for\ the}}$\\ $\underline{\it{intermolecular\ forces}}$ so
that the important issues such as the interplay between chemical
bondings and intermolecular forces can be
investigated~\cite{dalgarno,pople3}.




%


\vspace{0.45in}

\begin{thebibliography}{99}
\vspace{0.15in}
\bibitem{pauling} L. Pauling, The Nature of the Chemical Bond and
the Structure of Molecules and Crystals: An Introduction to Modern
Structure Chemistry, 3rd ed (Cornell University Press, 1960).
\bibitem{ruedenberg1}K. Ruedenberg, Rev. Mod. Phys. 34, 326 (1962).
\bibitem{lipscomb} W. N. Lipscomb, Boron Hydrides (W. A. Benjamin, 1963).
\bibitem{harris}D. A. Harris and M. D. Bertolucci, Symmetry and
Spectroscopy: An Introduction to Vibrational and Electronic
Spectroscopy (Oxford University Press; 1978).
\bibitem{hoffmann} R. Hoffmann, Rev. Mod. Phys. 60, 601 (1988).
 \bibitem{wei0} (a) L. Wei, Ph.D. Thesis, University of Illinois at
 Urbana-Champaign (UMI Publication, 1998); (b) private communication (1998).
\bibitem{kauzmann} W. Kauzmann, Quantum Chemistry: An Introduction
(Academic Press Inc., 1957).
\bibitem{cotton} F. A. Cotton, G. Wilkinson, C. A. Murillo and M.  Bochmann, Advanced Inorganic
Chemistry, 6th ed (John Wiley \& Sons, 1999).
\bibitem{slater}J. C. Slater, Quantum Theory of Atomic Structure, Vol. 1 (McGraw-Hill Book Company, Inc, 1960).
 \bibitem{kato}T. Kato, Perturbation Theory for Linear Operators (Springer: New York; 1980).
  \bibitem{marcus} R. A. Marcus, J. Phys. Chem. A 105, 2612 (2001).
 \bibitem{kirao} H. Kirao, Recent Advances in Multireference Methods (World Scientific, 1999).
   \bibitem{wei4} L. Wei, arXiv: physics/0412174 (2004).
\bibitem{pople1} (a) W. J. Hehre, L. Radom, P. v. R. Schleyer, and J. A. Pople, Ab Initio Molecular Orbital
 Theory (Wiley, New York, 1986); (b) V. A. Rassolov, M. A. Ratner, J. A. Pople, P. C. Redfern, and L. A. Curtiss,
 J. Comput. Chem. 22, 976 (2001).
\bibitem{dunning} T. H. Dunning, Jr., K. A. Peterson, and D. E. Woon, in Encyclopedia of
 Computational Chemistry, ed. P. v. R. Schleyer (John Wiley \& Sons: New York, 1998); pp. 88-115.
\bibitem{kestner} N. R. Kestner and J. E. Combariza, Rev. Comput. Chem. 13, 99 (1999).
 \bibitem{wei2} L. Wei, A. W. Jasper, and D. G. Truhlar, J. Phys. Chem. A 107,7236 (2003).
\bibitem{pople2} J. A. Pople, R. Krishnan, H. B. Schlegel, and J. S. Binkley,  Int. J. Quantum Chem. 14,
545 (1978).
\bibitem{karplus1} P. J. Rossky and M. Karplus, J. Chem. Phys. 72, 6085 (1980).
 \bibitem{bartlett} (a) R. J. Bartlett, Annu. Rev. Phys. Chem. 32, 359 (1981); (b)
 R. J. Bartlett, in Chemistry for the 21st Century, ed. E. Keinan and I. Schechter (John \& Wiley-VCH, 2001); pp. 271-286.
\bibitem{schaefer} T. D. Crawford, S. S. Wesolowski, E. F. Valeev, R. A. King, M. L. Leininger, and H. F. Schaefer,
 in Chemistry for the 21st Century, ed. E. Keinan and I. Schechter (John \& Wiley-VCH, 2001); pp. 219-246.
\bibitem{shavitt} I. Shavitt, in Method of Electronic Structure Theory, ed. H. F. Schaefer (Plenum Press, 1979);
 pp. 189-275.
\bibitem{goddard1} W. A. Goddard, T. H. Dunning, Jr., W. J. Hunt, and P. J. Hay, Acc. Chem. Res. 6, 368 (1973).
\bibitem{rudenberg2} K. Rudenberg and K. Sundberg in Quantum Science, ed. J. L. Calais, O.
Goscinski, J. Lindberg, and Y. $\ddot{O}$hrn (Plenum: New York,
1976); pp. 505-515.
\bibitem{peyerimhoff} R. J. Buenker and S. D. Peyerimhoff, in New Horizons of Quantum Chemistry,
ed. P.-O. L$\ddot{o}$wdin and B. Pullman (D. Reidel Publishing
Company, 1983);, pp. 183-219.
\bibitem{roos} (a) B. O. Roos, in Advances in Chemical Physics: Ab Initio Methods in Quantum Chemistry - II
 ed. Lawley, K. P. (John Wiley \& Sons Ltd.: Chichester, England, 1987); pp. 399-434;
(b) B. O. Roos, Acc. Chem. Res. 32, 137 (1999).
\bibitem{freed} X. C. Wang and K. F. Freed, J. Chem. Phys. 91, 3002 (1989).
 \bibitem{werner} H.-J. Werner, Mol. Phys. 89, 645 (1996).
 \bibitem{gordon} M. W. Schmidt and M. S. Gordon, Annu. Rev. Phys. Chem. 49, 233 (1998).
\bibitem{davidson1} E. R. Davidson and A. A. Jarzecki, in Recent Advances in
Multireference Methods, ed. H. Kirao (World Scientific, 1999); pp.
31-63.
 \bibitem{wei3} L. Wei, G. Li, and Y.-C. Chang, Surf. Sci. 439, 14 (1999).
 \bibitem{jortner} J. Jortner, S. A. Rice, and R. Silbey, in Modern Quantum Chemistry,
 Vol 3 (Academic Press, Inc.: New York, 1965); pp. 139-160.
\bibitem{fetter} Alexander L. Fetter and John Dirk Walecka, Quantum Theory of Many-Particle Systems (McGraw-Hill, 1971).
\bibitem{lindgren} I. Lindgren and J. Morrison, Atomic Many-Body
Theory (Springer-Verlag, 1982).
 \bibitem{kohn} (a) P. Hohenberg and W. Kohn, Phys. Rev. 136, B864 (1964); (b) W. Kohn and L. J.
Sham, Phys. Rev. 140, A1133 (1965).
 \bibitem{parr} R. Parr and W. Yang, Density-Functional Theory for
 Atoms and Molecules (Oxford University Press, 1989).
\bibitem{jursic} B. S. Jursic, in Recent Developments and Applications of Modern Density
Functional Theory, ed. J. M. Seminario (Elsevier: Amsterdam, 1996).
\bibitem{davidson2} E. R. Davidson, Int. J. Quantum Chem. 69, 241 (1998).
 \bibitem{tddft} K. Burke and E. K. U. Gross, in Density Functionals: Theory and Applications, Lecture Notes in Physics,
  Vol. 500, ed. D. Joubert (Springer: Heidelberg, 1998).
 \bibitem{szabo} A. Szabo and N. S. Ostlund, Modern Quantum
Chemistry: Introduction to Advanced Electronic Structure Theory
(McGraw-Hill, New York, 1989).
\bibitem{goddard2} C. F. Melius and G. A. Goddard, Phys. Rev. A 10, 1528 (1974).
\bibitem{hamann} (a) D. R. Hamann, M. Schluter, and C. Chiang, Phys. Rev. Lett. 43, 1494 (1979); (b) G. B. Bachelet,
 D. R. Hamann, and M. Schluter, Phys. Rev. B 26, 4199 (1982).
\bibitem{huzinaga} S. Huzinaga, L. Seijo, Z. Barandiaran, and M. Klobukowski, J. Chem. Phys. 86, 2132 (1987).
\bibitem{warshel} A. Warshel and M. Levitt, J. Mol. Biol. 103, 227 (1976).
\bibitem{karplus2}(a) M. J. Field, P. A. Bash, and M. Karplus, J. Comput. Chem. 11, 700 (1990);
(b) N. Reuter, A. Dejaegere, B. Maigret, and M. Karplus, J. Phys.
Chem. A 104, 1720 (2000).
 \bibitem{keiji} R. D. J. Froese and K. Morokuma, in Encyclopedia of Computational Chemistry, ed. P. v. R.
 Schleyer, N. L. Allinger, T. Clark, J. Gasteiger, P. A. Kollman, H. F. Schaefer III, P. R. Schreiner
 (John Wiley: Chichester,1998); pp. 1245-1257.
\bibitem{gao} J. Gao, P. Amara, C. Alhambra, and M. J. Field, J. Phys. Chem. A 102, 4714
(1998).
\bibitem{friesner} R. B. Murphy, D. M. Philipp, and R. A. Friesner, J. Comp. Chem. 21, 1442 (2000).
 \bibitem{jorgensen} W. L. Jorgensen, in Encyclopedia of Computational Chemistry,
 ed. P. v. R. Schleyer (John Wiley \& Sons: New York, 1998); pp. 1986-1989.
\bibitem{allinger} U. Burkert and N. L. Allinger, Molecular Mechanics, ACS Monograph 177, 1982.
 \bibitem{dalgarno} A. Dalgarno, Rev. Mod. Phys. 35, 611 (1963).
\bibitem{pople3} J. A. Pople, Faraday Discuss. Chem. Soc. 73, 7 (1982).
\end{thebibliography}
\end{document}